\documentstyle[twoside,fleqn,npb]{article}

\newcommand{\AmS}{{\protect\the\textfont2
  A\kern-.1667em\lower.5ex\hbox{M}\kern-.125emS}}

\title{Extraction of skewed parton distributions from experiment} 

\author{Andreas Freund
\address{I.N.F.N, Sezione Firenze, Largo Enrico Fermi 2, 50125 Firenze, Italy}}

\begin{document}

\begin{abstract}
In this paper we will discuss algorithms for extracting skewed parton 
distributions (SPD's) from experiment as well as the relevant process and experimental
observable suitable for the extraction procedure.
\end{abstract}

\maketitle

\section{Introduction}
\label{intro}

The basic concept of SPD's \cite{1,2,3,4,5} is best 
illustrated with the lowest order graph of deeply virtual Compton scattering
(DVCS) in which a quark of momentum 
fraction $x_1$ leaves the proton and is returned to it with $x_2$. 
The two fractions not being equal is due to the fact 
that an on-shell photon is produced which necessitates a change in the $+$ 
momentum in going from the virtual space-like photon with $+$ momentum $-x_{bj}p_+$,
to basically zero $+$ momentum of the 
real $\gamma$. This sets $x_2=x_1-x$ and thus the skewedness parameter to $x$.
Thus one has a nonzero momentum transfer onto the proton and the parton 
distributions (PDF's) which enter the process are non longer the regular PDF's since 
the matrix element of the
quark (gluon) operator is now taken between states of unequal momentum rather
than equal momentum.

\section{Appropriate Process and experimental observable}
\label{proc}

The most desirable process for extracting SPD's is the one
with the least theoretical uncertainty, the least singular $Q^2$ behavior and 
a proven factorization formula. 

The process which fulfills all the above criteria is DVCS and
the experimental observable which allows direct access to the SPD's
is the azimuthal angle asymmetry $A$ of the combined DVCS and 
Bethe-Heitler(BH) differential cross section. 
$A$ is defined as \cite{3}:
\begin{equation}
A =\frac{\int^{\pi/2}_{-\pi/2}d\phi~d\sigma - \int^{3\pi/2}_{\pi/2}
    d\phi~d\sigma}{\int^{2\pi}_{0}d\phi~d\sigma}.
\label{asym}
\end{equation}

The reason why this asymmetry is not $0$ is due to the interference term between
BH and DVCS which is proportional to the real part of the DVCS amplitude. 
The factorized expression for the real part of the amplitude is \cite{2} 
\begin{eqnarray}
Re~T(x,Q^2) &=&  \int^{1}_{-1+x}\frac{dy}{y} Re~C_{i}(x/y,Q^2)\nonumber\\
& & f_{i}(y,x,Q^2).
\label{factor}
\end{eqnarray}
$Re~C_{i}$ is the real part of the hard scattering coefficient (HSC) and $f_i$ are the
SPD's.  
At HERA one is 
mainly restricted to the small-$x$ region where gluons dominate and thus $i$ will be $g$. 
Thus Eq.\ (\ref{asym}) contains only measurable or directly computable quantities
except Eq.\ (\ref{factor}) in the interference part.
Therefore, one would now be
able to extract the SPD's from 
on $A$, if one could deconvolute Eq.\ (\ref{factor}). 
However, for SPD's life is not as "simple" as in inclusive DIS where the 
deconvolution for $F_2$ is trivial, since the
SPD's depend on two rather than one variable. Furthermore,
the HSC depends on the same variables as the SPD's. 
These facts make the deconvolution of Eq.\ (\ref{factor}) impossible.

\section{Algorithms for extracting SPD's}
\label{extract}

Rather than deconvoluting, one can expand the PDF's with respect to a 
complete set of orthogonal polynomials 
$P^{(\alpha_P)}_{j}(t)$. In this particular case we need the orthogonality of
the polynomials to be on the interval $-1 \leq t \leq 1$ with 
$t=\frac{2y-x}{2-x}$ equivalent to 
$-1+x\leq y \leq 1$ as found in Eq.\ (\ref{factor}). One can then write the following expansion:
\begin{eqnarray}
f^{q,g}(y,x,Q^2) &=& \frac{2}{2-x}\sum^{\infty}_{j=0}
\frac{w(t|\alpha_P)}{n_j(\alpha_P)}P_{j}^{q,g}(t)\nonumber\\
& & M^{q,g}_j(x,Q^2)
\label{expand}
\end{eqnarray}
with $w(t|\alpha_P)$, $n_j(\alpha_P)$ and $\alpha_P$ being weight, normalization 
and a label determined by the choice of the orthogonal polynomial used. 
$M^{q,g}_j(x,Q^2)$ is given by:
\begin{equation}
M^{q,g}_j(x,Q^2) = \sum^{\infty}_{k=0} E^{q,g}_{jk}(x)f^{q,g}_k(x,Q^2),
\label{coeff}
\end{equation}
where
\begin{equation}
f^{q,g}_j(x,Q^2) = \sum^{j}_{k=0} x^{j-k}B^{q,g}_{jk}\tilde f^{q,g}_k(x,Q^2).
\end{equation}
$B^{q,g}_{jk}$ is an operator transformation matrix which fixes the NLO 
corrections to the eigenfunctions of the kernels and is thus just the identity 
matrix in LO. 
The upper limit in Eq.\ (\ref{coeff}) is given by the constraint 
$\theta$-functions
present in the expansion coefficients, which are generally defined by
\begin{eqnarray}
& &E_{jk}(\nu;\alpha_P|x) =\frac{\theta_{jk}}{(2x)^k}\frac{\Gamma (\nu) \Gamma (\nu + k)}
{\Gamma (\frac{1}{2}) \Gamma (k + \nu + \frac{1}{2})}\nonumber\\
& & \int_{-1}^1 dt (1-t^2)^
{k+ \nu - \frac{1}{2}}
\frac{d^k}{dt^k}P_j^{\alpha_P}\left ( \frac{xt}{2-x} \right ).
\end{eqnarray}
The moments of the SPD's evolve according to
\begin{equation}
\tilde f^{q,g}_j(x,Q^2) = \tilde E_j (Q^2,Q^2_0)
\tilde f^{q,g}_j(x,Q^2_0)
\end{equation}
where the evolution operator is a matrix of functions in the singlet case.
Finally, the Gegenbauer moments of the SPD's at $Q_0^2$ 
are defined by
\begin{eqnarray}
\tilde f^{q}_{j}(x,Q_0^2) &=& \int^{1}_{-1}dt~\left ( \frac{x}{2-x} \right )^j
\nonumber\\
& & C_j^{3/2}\left ( \frac{tx}{2-x} \right ) f^{q}(t,x,Q^2_0)\nonumber\\
\tilde f^{g}_{j}(x,Q_0^2) &=& \int^{1}_{-1}dt~\left ( \frac{x}{2-x}\right )^{j-1}
\nonumber\\
& &C_{j-1}^{5/2}\left( \frac{tx}{2-x} \right ) f^{g}(t,x,Q^2_0).
\label{moments}
\end{eqnarray} 

In LO order and at small x the above formalism simplifies.
Owing to the conformal properties of the operators involved in the definition
of the SPD's one finds the following expansion
\begin{eqnarray}
f^g(y_1,x,Q^2)&=& \frac{2}{2-x}\sum^{\infty}_{j=0}\sum^{\infty}_{k=1}
\frac{w(t|5/2)}{N_j(5/2)}\nonumber\\
& &E^{g}_{jk-1}(x)C_{j-1}^{5/2}(t)\tilde f^g_{k-1}(x,Q^2)\nonumber\\
f^q(y_1,x,Q^2) &=& \frac{2}{2-x}\sum^{\infty}_{j=0}\sum^{\infty}_{k=0}
\frac{w(t|3/2)}{N_j(3/2)}\nonumber\\
& &E^{q}_{jk}(x)C_j^{3/2}(t)\tilde f^q_k(x,Q^2),
\label{expansion}
\end{eqnarray}
with $w(t|\nu) = (t(1-t))^{\nu -1/2}$ and the $C_j^{\nu}$'s being Gegenbauer 
polynomials .
The multiplicatively renormalizable moments evolve as above but with the 
explicit evolution operator:
\begin{equation}
\tilde E^{ik}_j(Q^2,Q^2_0) = T e^{\left( -\frac{1}{2}
\int^{Q^2}_{Q^2_0}\frac{d\tau}{\tau}\gamma^{ik}_j(\alpha_s(\tau)) \right )}
\end{equation}
where $T$ orders the matrices of the regular LO anomalous dimensions (i,k = q,g) along the 
integration path.

Inserting Eq.\ (\ref{expansion}) in Eq.\ (\ref{factor}) one obtains:
\begin{eqnarray}
& & Re~T(x,Q^2) = 2\sum^{\infty}_{j=0}\sum^{\infty}_{k=1}
\tilde E_{k-1} (Q^2,Q^2_0)\nonumber\\
& & \tilde f^{g}_{k-1}(x,Q^2_0)E^{g}_{jk-1}(x)\int^{1}_{-1}\frac{dt}{2t+x}
\frac{w(t|5/2)}{N_j(5/2)}\nonumber\\
& & Re~C_{g}\left ( \frac{1}{2}+\frac{t}{x},Q^2\right )C_{j-1}^{5/2}(t),
\label{mastereq}
\end{eqnarray}
where we chose the factorization/renormalization scale to be equal to $Q^2$.
As one can see the integral in the sum is now only over known functions and 
will yield, for fixed $x$, a function of $j$ as will also do the expansion 
coefficients for fixed $x$. The evolution operator can also be evaluated and will yield 
for fixed $Q^2$ also just a function of $j$, which 
leaves the coefficients $\tilde f^{g}_{k-1}(x,Q^2_0)$ as the only unknowns. 
Since the lefthand side will be known from experiment
for fixed $x$ and $Q^2$, we are still in the unfortunate situation that a 
number is determined by the sum over $j$ of an infinite number of coefficients.
However measuring the real part at an infinite number of $Q^2$ for fixed $x$, one would have an 
infinite 
dimensional column vector on the lefthand side and on the right hand 
side one would have a square matrix times another column vector of 
coefficients of which the dimension is determined by the number of $j$. Since 
all the entries in the matrix are real and positive definite provided that there
are no zero eigenvalues one can find the inverse. Thus one can 
directly compute the moments of our initial parton distributions which are needed to reconstruct the
skewed gluon distribution from Eq.\ (\ref{expand}).  

The drawback of the above procedure is that this process has to be repeated 
anew for each $x$ and that NLO evolution studies \cite{4} indicate that one might 
need as amuch as $50-100$ polynomials to achieve enough accuracy at small-$x$.
This, of course, would render this procedure useless in an experimental situation
where a $j$ of $5-10$ is possibly achievable!
 
A practical way out of the above mentioned predicament is by making a simple
minded ansatz for the skewed gluon distribution in the different regions like
$A_0z^{-A_1}(1-z)^A_3$ for the DGLAP region where $z$ is now just a dummy 
variable, plug this form in Eq.\ (\ref{factor}) and fit the coefficients to the
data of the real part of the DVCS amplitude for fixed $x$ and $Q^2$. 
One can repeat this procedure for different values of $Q^2$ and then interpolate
between the different coefficients to obtain a functional form of the 
coefficients in $Q^2$, alternatively, after having extracted the values of
the coefficients for different values of $x$ at the same $Q^2$, use an evolution
programm with the ansatz and the fitted coefficients as input and check whether
one can reproduce tha data for the real part at higher $Q^2$, thus checking
the viability of the model ansatz.

To obtain an ansatz fullfilling the various constraints for SPD's 
(see Ji's and Radyushkin's references in \cite{1}), one should start from
the double distributions (DD) (see Redyushkin's 
references in \cite{1}.) which yield the skewed gluon distribution
in the various regions
\begin{eqnarray}
g(y,x) &=& \theta (y \geq x)\int^{\frac{1-y}{1-x}}_0 dz G(y-xz,z) +\nonumber\\
& & \theta (y \leq x) \int^{\frac{y}{x}}_0 dz G(y-xz,z).
\end{eqnarray}
Due to the fact that there are no anti-gluons, the above formula is enough to 
cover the whole region of interest $-1+x\leq y leq 1$. What remains is to choose
an appropriate model ansatz for G, for example,
\begin{equation}
G(z_1,z) = \frac{h(z_1,z)}{h(z_1)}f(z_1)
\end{equation}
with $f(z_1)$ being taken from a diagonal parametrization with its 
coefficients now being left as variants in the skewed case and
the normalization condition $h(z_1) = \int^{1-z1}_0 dz h(z_1,z)$ such that, in the
diagonal limit, the DD just gives the diagonal distribution. The choice for 
$h(z_1,z)$ is a matter of taste but should be kept as simple as possible.

\section{Conclusions and outlook}
\label{concl}

After having showed, that the extraction of skewed parton distributions from
DVCS experiments is principally as well as practically possible.

\section*{Acknowledgments}

This work was supported by the E.\ U.\ contract $\#$FMRX-CT98-0194.

\end{document}